\documentclass[onecolumn,showpacs,superscriptaddress,amsmath,amssymb]{revtex4}

\usepackage{graphicx}
\usepackage{dcolumn}
\usepackage{bm}
\usepackage{booktabs}
\usepackage{color}
\usepackage{mathrsfs}
\usepackage{ulem}
\usepackage{slashbox}
\usepackage[bf,raggedright,indentafter]{titlesec}

\begin{document}\large
\title{ Generalized susceptibilities along the phase boundary of the three-dimensional, three-state Potts model}

\author{Xue Pan}\email{panxuepx@gmail.com}
\affiliation{Key Laboratory of Quark and Lepton Physics (MOE) and
Institute of Particle Physics, Central China Normal University, Wuhan 430079, China}
\author{Mingmei Xu}
\affiliation{Key Laboratory of Quark and Lepton Physics (MOE) and
Institute of Particle Physics, Central China Normal University, Wuhan 430079, China}
\author{Yuanfang Wu}\email{wuyf@phy.ccnu.edu.cn}
\affiliation{Key Laboratory of Quark and Lepton Physics (MOE) and
Institute of Particle Physics, Central China Normal University, Wuhan 430079, China}

\begin{abstract}

 Through the Monte Carlo simulation of the three-dimensional, three-state Potts model, which is a paradigm of finite-temperature pure gauge QCD, we study the fluctuations of generalized susceptibilities near the temperatures of external fields of first-, second-order phase transitions and crossover. Similar peak-like fluctuation appears in the second order susceptibility at three given external fields. Oscillation-like fluctuation appears in the third and fourth order susceptibilities. We find that these non-monotonic fluctuations are not only associated with the second-order phase transition, but also the first-order one and crossover in a system of finite-size. We further present the finite-size scaling analysis of the second and fourth order susceptibilities, respectively. The exponent of the scaling characterizes the order of the transitions, or the crossover.

\end{abstract}



\maketitle

\section{Introduction}

It has been shown that the new form of matter---quark gluon plasma has been produced at the Relativistic Heavy-Ion Collider (RHIC)~\cite{nature-QGP}. However, the phase boundary is not yet clear in the plane of temperature-baryon chemical potential ($T-\mu_B$). One of the main goals of the Beam Energy Scan (BES) program at RHIC is to map the Quantum Chromodynamics (QCD) phase diagram.

It's expected that the system undergoes a first-order phase transition at high baryon chemical potential and low temperature~\cite{Potts-QCD 1, confinement first 1, confinement first 2, confinement first 3, confinement first 4, confinement first 5}. With decrease of baryon chemical potential and increase of temperature, the first-order phase transition line ends at a critical point, which belongs to the three-dimensional Ising universality class~\cite{cp1, cp2, Stephanov-prl81}. At vanishing chemical potential, the calculations of lattice QCD have shown it is a crossover~\cite{nature-crossover, hotQCD-crossover}, see a sketch of the QCD phase diagram in figure 1(a).

To map the QCD phase diagram, the high-order cumulants of conserved charges are proposed as sensitive observable of the second-order phase transition~\cite{koch}. They are related to the generalized susceptibilities and can be calculated by lattice QCD near vanishing chemical potential, i.e., the position of the blue arrow line as showed in figure 1(a), where the transition appears as the remnants of a second-order phase transition belonging to the three-dimensional $O(4)$ universality class~\cite{MCheng-PRD79, EPJC67, Vladi, WJF-PRD81, BFriman-EPJC}. At the critical point
of finite chemical potential, the position of the red arrow line as showed in figure 1(a), effective models of QCD
have also calculated the cumulants of conserved charges~\cite{NPA, Asakawa-prl103, Stephanov-prl107}. Both of them show that non-monotonic fluctuations, e.g. peak structure and oscillation, of generalized susceptibilities are signatures of the critical point~\cite{MCheng-PRD79, Vladi, WJF-PRD81, BFriman-EPJC, NPA, Asakawa-prl103, Stephanov-prl107}.

It is interesting to ask if the system crosses the phase boundary at other places, such as the region of the first-order phase transition, e.g., the black arrow line as shown in figure 1(a), how does the behavior of the generalized susceptibilities change? A model investigation should be helpful.

The three-dimensional, three-state Potts model is one of the standard paradigms of finite-temperature pure gauge QCD. It has the same $Z(3)$ global symmetry as QCD system~\cite{second order 1, Potts-QCD 1, Potts-QCD 2, Potts-QCD 3, Potts-QCD 4}. Its phase boundary in temperature and external field plane is shown in figure 1(b). It is comparable to that of the QCD. At vanishing external field, the temperature-driven phase transition has proved to be of first-order~\cite{first order 1, first order 2}. With increase of the external field, the first-order phase transition weakens and ends at a critical point $(1/T_c, h_c) = (0.54938(2), 0.000775(10))$, which also belongs to the three-dimensional Ising universality class~\cite{second order 1, second order 2}. Beyond the critical point, it is a crossover region. Both transitions of QCD deconfinement and QCD chiral symmetry restoration have the line of first-order phase transition, and end in a second-order endpoint. The baryon chemical potential just acts as the external field in the Potts model.

In this paper, we will first show the fluctuations of the second to fourth order susceptibilities in the vicinity of the phase transition temperatures at three given external fields of first-, second-order phase transitions and crossover of the Potts model, i.e., cross the phase boundary along the black, red, blue arrow lines of figure 1(b).

The paper is organized as follows. The corresponding susceptibilities of the magnetization from the three-dimensional, three-state Potts model is firstly derived in section 2. The second to the fourth order susceptibilities at different system sizes of external fields of first-, second-order phase transitions and crossover are presented in section 3. It is noteworthy that they all exist non-monotonic fluctuations. The finite-size scaling of the second and fourth order susceptibilities are further analyzed in section 4. It shows that the scaling exponents of them change gradually with the external field. Thus it can be used to distinguish the order of the transition. Finally, the summary and conclusions are given in section 5.

\begin{figure*}[htb]
\begin{center}
\includegraphics[width=0.45\textwidth,angle=0]{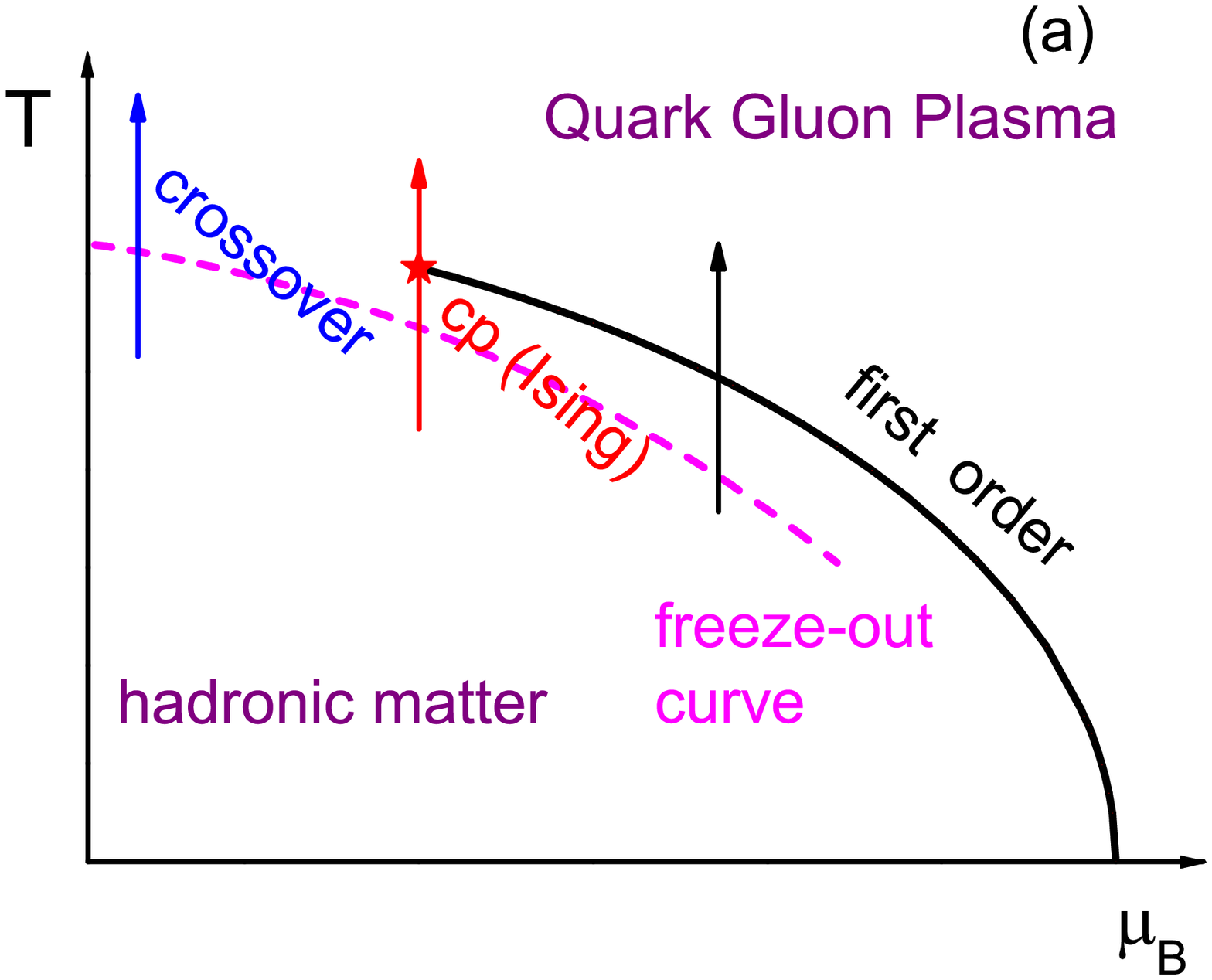}
\includegraphics[width=0.45\textwidth,angle=0]{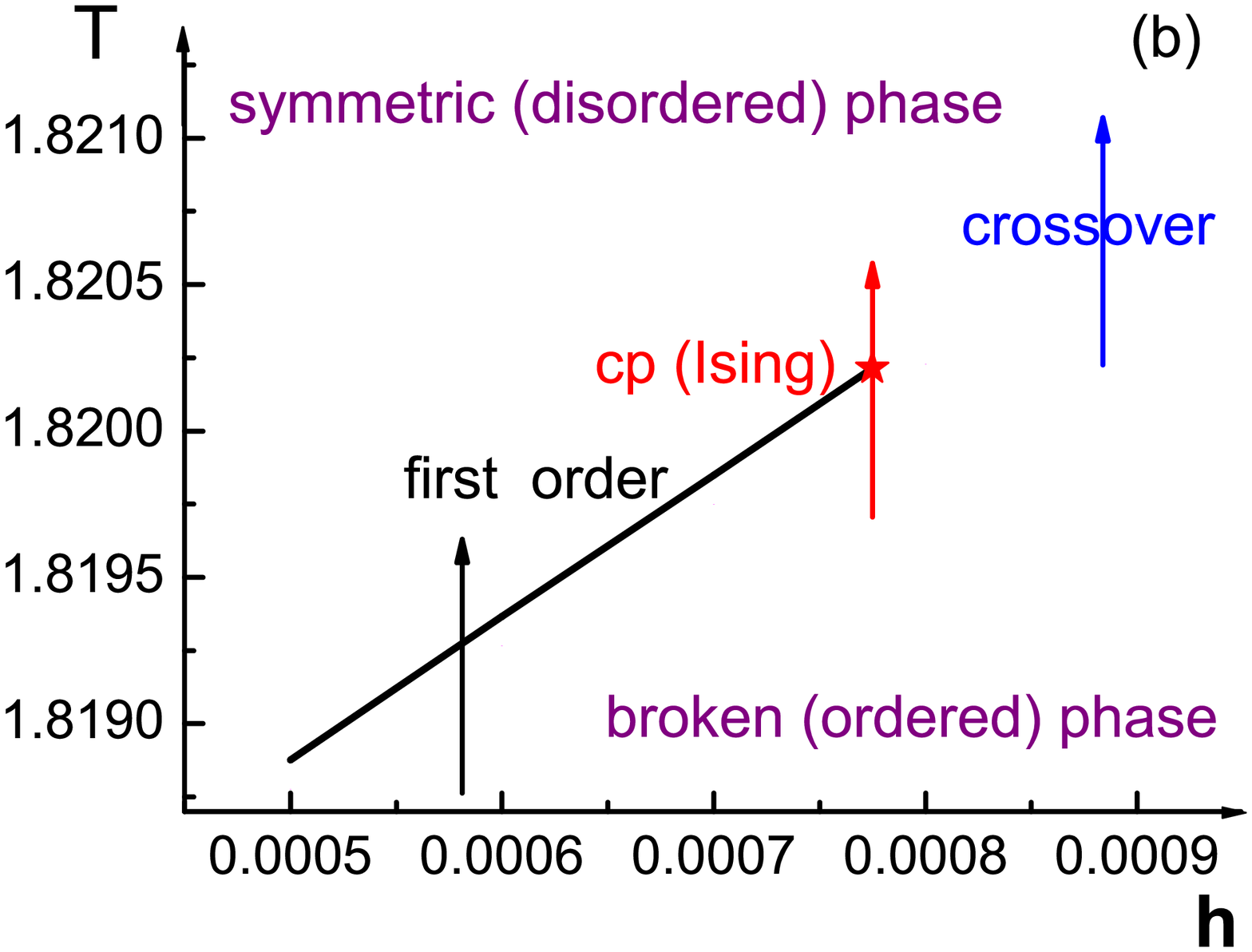}
\caption{\label{Fig. 1}(Color online) a sketch of the QCD phase diagram with the freeze-out curve on the temperature-baryon chemical potential plane (a) and the phase diagram of the three-dimensional, three-state Potts model on the temperature-external field plane (b). }
\end{center}
\end{figure*}

\section{Generalized susceptibilities in the three-dimensional, three-state Potts model}

The three-dimensional three-state Potts model is described in terms of spin variable $s_i \in\{1,2,3\}$, which is located at sites $i$ of a cubic lattice of size $V=L^3$. The partition function of the model is defined by,
\begin{flalign}\label{partition func}
&\qquad Z(\beta, h) =\sum_{\{{s}_{i}\}}e^{-(\beta E-hM)}.&
\end{flalign}
Where $\beta=1/T$, and $h=\beta H$ is normalized external magnetic field. $E$ and $M$ are energy and magnetization, respectively,
\begin{flalign}\label{energy}
&\qquad E=-J\sum_{\langle i,j\rangle}\delta(s_i,s_j), ~~~~~~~M=\sum_{i}\delta(s_i,s_g).&
\end{flalign}
Here $J$ is an interaction energy between nearest-neighbour spins $\langle i,j\rangle$. In this work, $J$ is set as 1. $s_g$ is the direction of the ghost spin. A non-vanishing external field $h>0$ favours magnetization in this direction.

The order parameter is defined by the mean of magnetization~\cite{FY Wu},
\begin{flalign}\label{order parameter}
&\qquad m = \frac{3}{2}\frac{\langle M\rangle}{V}-\frac{1}{2}.&
\end{flalign}
While the order parameter in QCD pure gauge theories is presented by Polyakov loop, which plays the role of magnetization. It signals the spontaneous breaking of center symmetry in the deconfined phase.

The susceptibility $\chi_2$ is the second order derivative of free energy density $f=-\frac{1}{V}\ln Z$ with respect to the external field,
\begin{flalign}\label{susceptibility}
&\qquad \left. \chi_2=-(\frac{\partial^2 f}{\partial h^2})\right |_T=\frac{1}{V}(\langle M^2\rangle-\langle M\rangle^2).&
\end{flalign}
Without $1/V$, it is the second order cumulant of the magnetization. Then the third and fourth order generalized susceptibilities are as follows, i.e.,
\begin{flalign}\label{cumulants of order parameter}
&\qquad \chi_3=\frac{1}{V}{\langle \delta {M}^3 \rangle},& \nonumber \\
&\qquad \chi_4=\frac{1}{V}(\langle \delta {M}^4 \rangle-3\langle\delta {M}^2 \rangle^2),&
\end{flalign}
where $\delta {M}=M-\langle M\rangle$.

In the following section, we will present and discuss the fluctuations of generalized susceptibilities at various system sizes at three external fields of the first-, second-order phase transitions and crossover of the three-dimensional, three-state Potts model.

\section{Generalized susceptibilities nearby the phase boundary of the three-dimensional, three-state Potts model}

In the three-dimensional, three-state Potts model, we first choose three fixed external fields. They are located in the region of first-, second-order phase transitions and crossover, respectively. i.e., very small external field $h = 0.0005$ for the first-order phase transition, $h=0.000775$ for the critical point~\cite{second order 2}, and $h=0.002$ for the crossover. Their positions in figure 1(b) are marked respectively by black, red and blue arrows.

For each external field, we performed simulations for 3 to 6 $\beta$-values. For each pair of couplings $(\beta, h)$, we generated 50000 independent configurations. They have been used in a Ferrenberg-Swendsen reweighting analysis~\cite{multi-histogram} to calculate observables at intermediate parameter values. Where the simulation is performed by the Wolff cluster algorithm~\cite{wolff}, and the helical boundary conditions are used. For each case, we carry out the simulation at various of lattice sizes of $L^3$ from $L$ = 40 to 70.

The second order susceptibility $\chi_2$ near the phase transition temperature at three external fields, $h$ = 0.0005, 0.000775, and 0.002, are presented in figure 2(a), 2(b) and 2(c), respectively. It is shown that at the first- (figure 2(a)), second-order phase transitions (figure 2(b)) and crossover (figure 2(c)), $\chi_2$ all exhibit a peak structure. As we know, for a first-order phase transition, the susceptibility is a $\delta$-function in an infinite system. It is divergent at a second-order phase transition. Finite system size makes it to be a finite peak in both types of transitions~\cite{Imry,Murry}.

With increase of system size, the position of the peak shifts to the higher temperature. In each sub-figure $T_c$ is set by the position of the maximum at the largest system size $L = 70$. In figure 2(a), 2(b) and 2(c), they are 0.54979, 0.549385, and 0.54769, respectively.

\begin{figure*}[htb]
\begin{center}
\includegraphics[width=0.9\textwidth,angle=0]{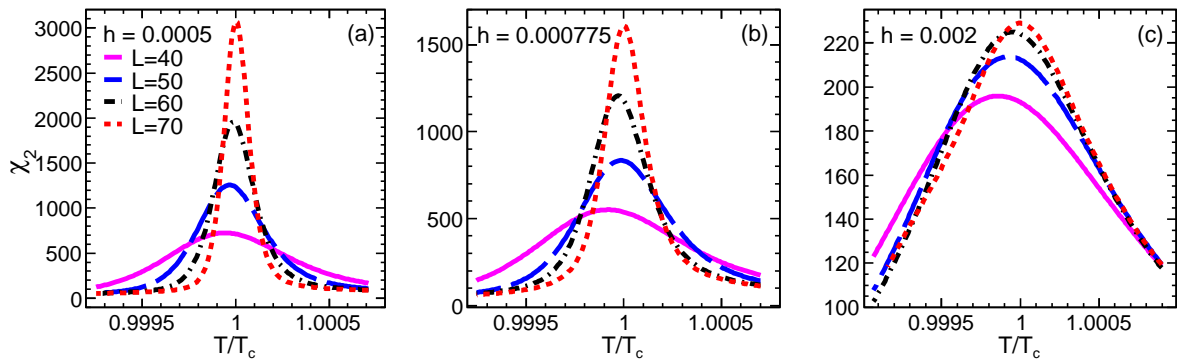}
\caption{\label{figure 2}(Color online) $\chi_2$ versus $T/T_c$ at $h$ = 0.0005 (a), 0.000775 (b) and 0.002 (c) with the system sizes of $L$ = 40, 50 , 60, and 70.}
\end{center}
\end{figure*}

The difference among figure 2(a), 2(b) and 2(c) is the height and width of the peak. From figure 2(a) to 2(c), the height of peak decreases but width becomes wider with the increase of the external field. Obviously, the peaks in figure 2(a) and 2(b) are narrower and sharper than those in figure 2(c). This implies although the structure of $\chi_2$ is similar in three different phase transition regions, it's more pronounced at the first- and second-order phase transitions than in the crossover region.

In a given sub-figure, the peak gets higher and sharper with increase of system size. It is clear that the size dependence becomes weaker and weaker with the increase of the external field, in particular in figure 2(c), the peaks at different system sizes are very close to each other.

The third order susceptibility $\chi_3$ near the phase transition temperature at three external fields, $h$ = 0.0005, 0.000775, and 0.002 are shown in figure 3(a), 3(b) and 3(c), respectively. It is obvious that the basic structure of each curve in three sub-figures is similar. It oscillates from negative to positive in the vicinity of the phase transition temperature. The third order susceptibility fluctuates more frequently and violently near the phase transition temperature than that of the second order one. Their dependence on the external field and system size are the same as those of the second order susceptibility.

\begin{figure*}[htb]
\begin{center}
\includegraphics[width=0.9\textwidth,angle=0]{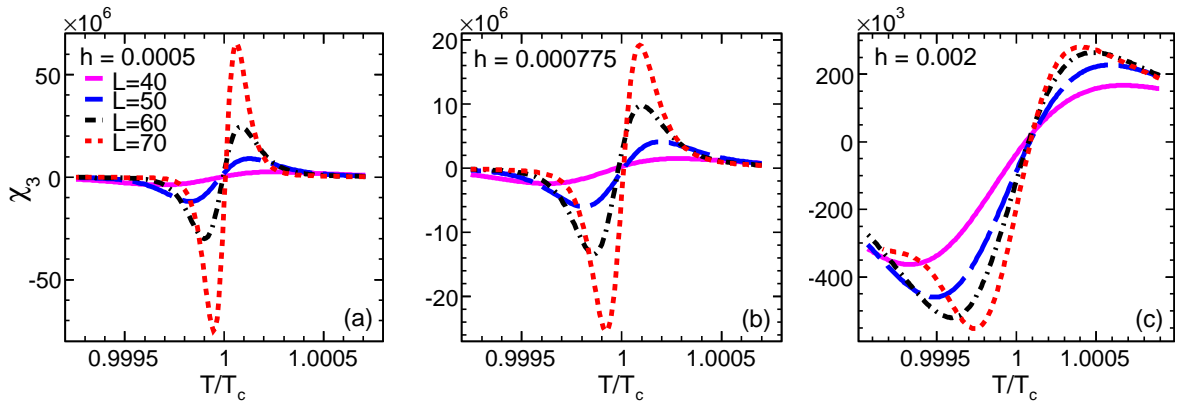}
\caption{\label{figure 3}(Color online) $\chi_3$ versus $T/T_c$ at $h$ = 0.0005 (a), 0.000775 (b) and 0.002 (c) with the system sizes of $L$ = 40, 50 , 60, and 70.}
\end{center}
\end{figure*}

The fourth order susceptibility $\chi_4$ is presented in figure 4. At $h=0.0005$ (figure 4(a)) and $h=0.000775$ (figure 4(b)), it has two positive peaks at the two sides of the phase transition temperature and between them is a negative valley. While, such structure is not fully shown up at $h=0.002$ (figure 4(c)). It shows again that the fluctuations in the crossover region are more smooth than those at the first- and second-order phase transition regions. The structure of $\chi_4$ is more complicated than $\chi_3$.

\begin{figure*}[htb]
\begin{center}
\includegraphics[width=0.9\textwidth,angle=0]{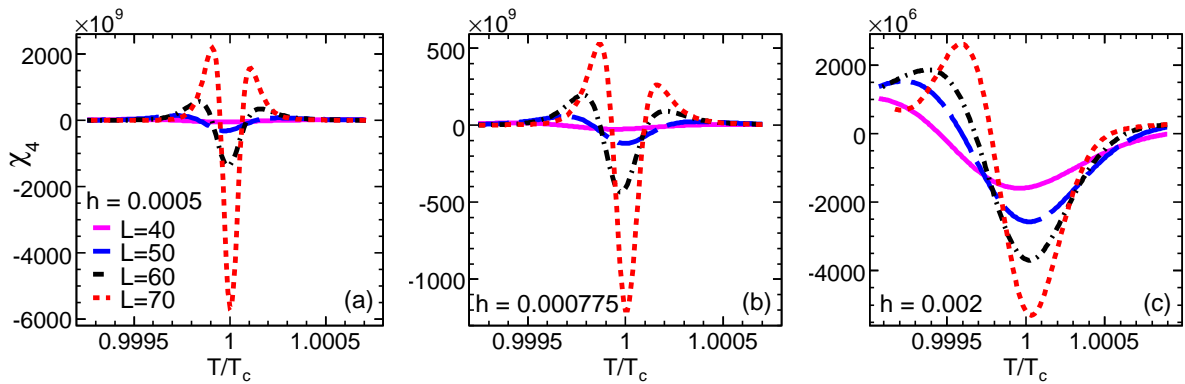}
\caption{\label{figure 4}(Color online) $\chi_4$ versus $T/T_c$ at $h$ = 0.0005 (a), 0.000775 (b) and 0.002 (c) with the system sizes of $L$ = 40, 50 , 60, and 70.}
\end{center}
\end{figure*}

In figure 2, figure 3, and figure 4, it is clear that the generic structure of the second, third and fourth order susceptibilities in the vicinity of the phase transition temperature is qualitatively similar at the three external fields. A peak appears at the second order susceptibility, oscillation and sign change happen at the third and fourth order susceptibilities. This shows that non-monotonic behavior, or a sign change is not only related to the critical fluctuations, but also the fluctuations at a first-order phase transition and crossover in a system of finite-size.

However, quantitatively, at an external field of crossover, the fluctuation changes with temperature obviously more smooth than that at the first- and second-order phase transitions. Moreover, all generalized susceptibilities are much weakly dependent on the system size. In fact, system-size dependence on the external field becomes weak gradually with the increase of the external field from sub-figure (a) to (c) in figure 2, figure 3, and figure 4. It is interesting to see the finite-size scaling behavior of the generalized susceptibilities at the first-, second-order phase transitions and crossover.

\section{Finite-size scaling of generalized susceptibilities}

As we know, susceptibility ($\chi_2$) satisfies the finite-size scaling relation at first- and second-order phase transitions. In the former case, it is determined only by the geometric dimension ($d$) of the system, and the height ($\chi_2^{max}$) is proportional to the volume $V$~\cite{Murry, Binder-review, Binder-first, RG-first}. While in the latter case, the height is approximately in proportion to a power of $V$, which is determined by the critical exponents. For an analytic crossover, the height of the susceptibility is independent of the system size when system size is large enough~\cite{nature-crossover}. The specific expression of $\chi_2^{max}$ to the size $L$ can be given as follows,
\begin{flalign}\label{fss chi2}
&\qquad \chi_2^{max}(L) \propto L^{\lambda_{\chi_2}},&
\end{flalign}
where the scaling exponent $\lambda_{\chi_2}$ equals to $d$ and 0 for a first-order phase transition and crossover, respectively. For a second-order phase transition, it is usually a fraction between $d$ and 0, which is decided by the universality class. So along the phase boundary of the three-dimensional, three-state Potts model, i.e., from a first-order phase transition to a second-order one, then to crossover, the scaling exponent keeps decreasing from $d$ to 0.

The logarithm of equation \eqref{fss chi2} can be expressed as,
\begin{flalign}\label{ln fss chi2}
&\qquad \ln{\chi_2^{max}} = \lambda_{\chi_2} \ln L + C_1,&
\end{flalign}
where $C_1$ is a constant. This is a straight line with respect to $\ln L$, whose slope is just the scaling exponent $\lambda_{\chi_2}$.

We plot in figure 5(a) the logarithm $\ln\chi_2^{max}$ as a function of $\ln L$ at various system sizes for $h = 0.0005$ (solid black circles), 0.000775 (solid red stars), and 0.002 (solid blue down-triangles). It is demonstrated that the same color points are well fitted by a straight line. So the height of the susceptibility at a given external field of the three-dimensional, three-state Potts model is well described by finite-size scaling, i.e., equation \eqref{fss chi2}, or equation \eqref{ln fss chi2}.

\begin{figure}[htb]
\includegraphics[width=0.75\textwidth,angle=0]{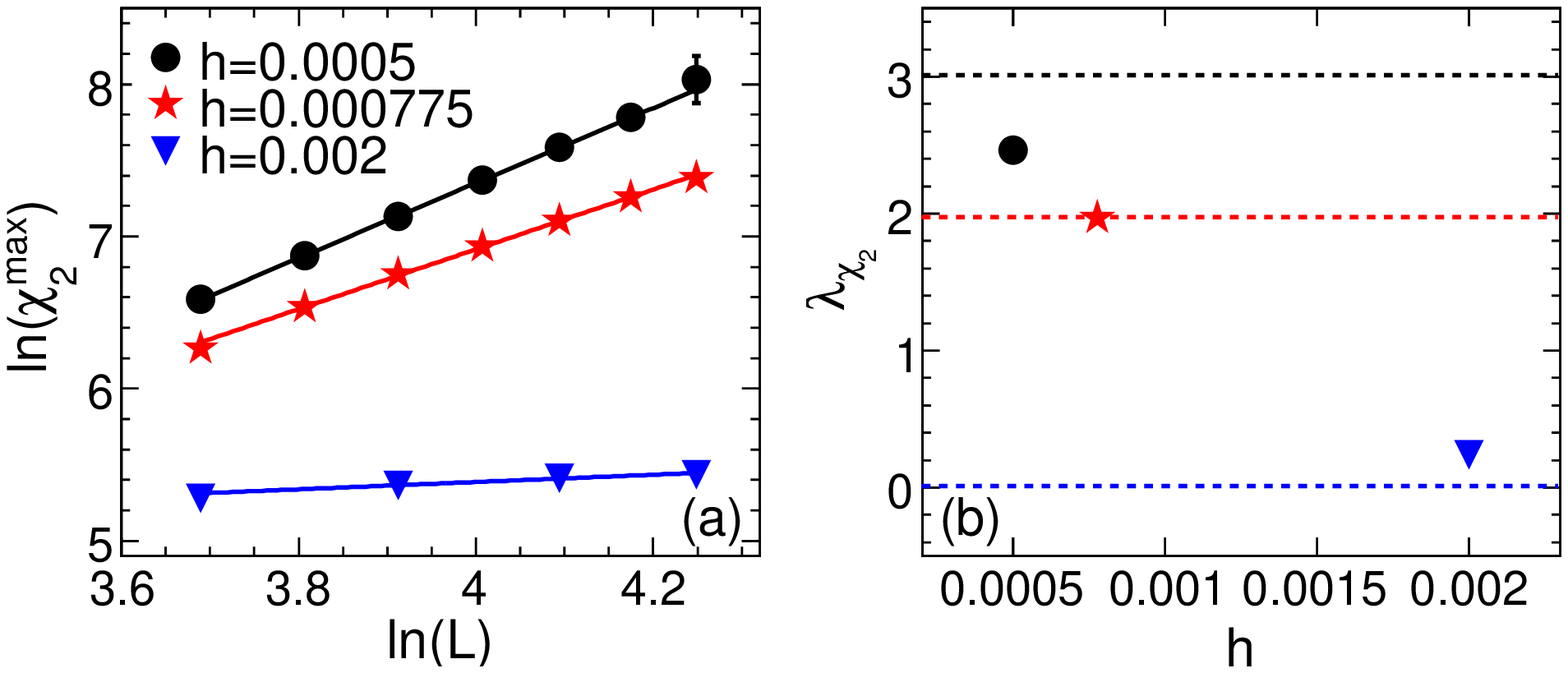}
\includegraphics[width=0.75\textwidth,angle=0]{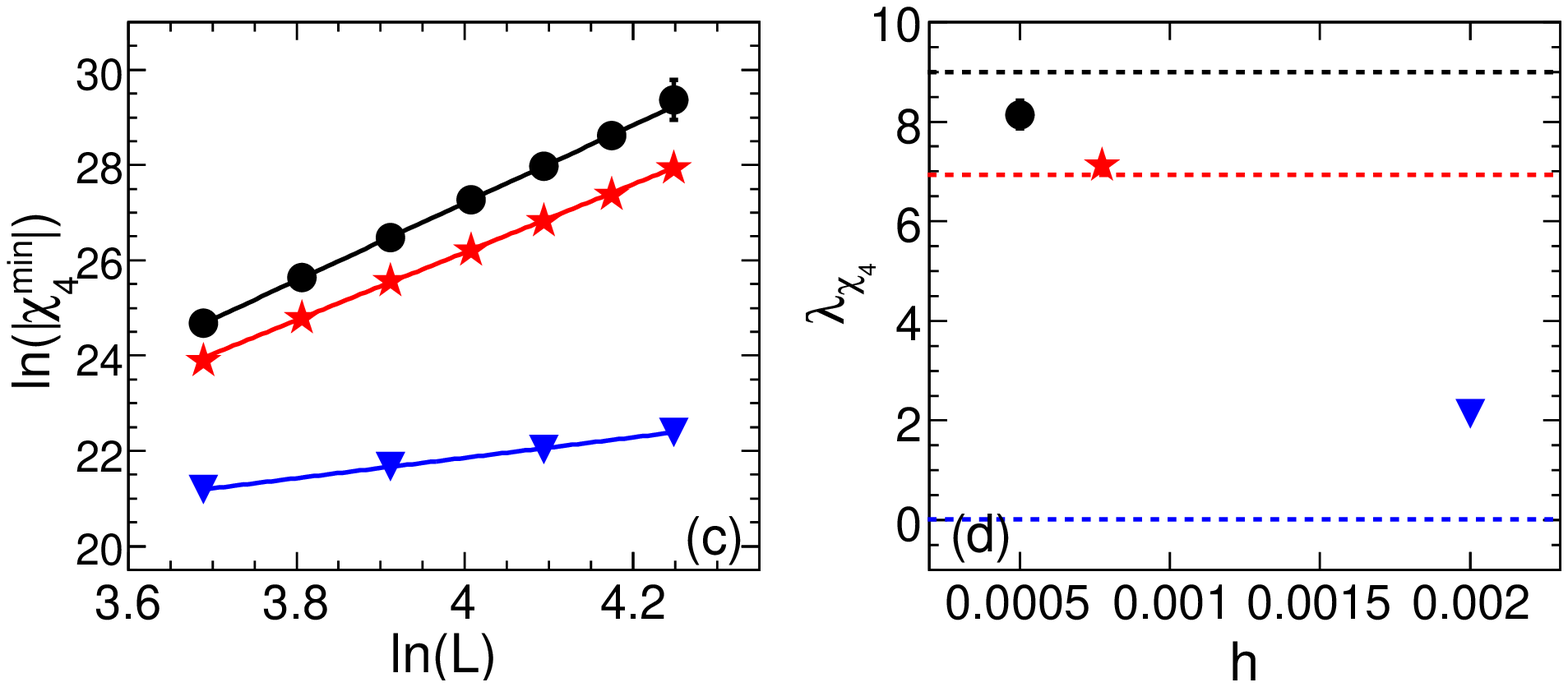}
\caption{\label{figure 5}(Color online). The logarithm of the hight for the second order susceptibility $\ln (\chi_2^{max})$  versus the logarithm of the system size $\ln (L)$ (a) and the values of the slope (b) for different external fields. The logarithm of the depth for the fourth order susceptibility $\ln (|\chi_4^{min}|)$  versus the logarithm of the system size $\ln (L)$ (c) and the values of the slope (d) for different external fields. The black (up), red (middle), and blue (down) dash lines show the expected values of $\lambda_{\chi_2}$ (b) and $\lambda_{\chi_4}$ (d) for a first-order phase transition, a second-order phase transition belonging to the three-dimensional Ising universality class and crossover, respectively.}
\end{figure}

The slopes of three solid lines ($\lambda_{\chi_2}$) in figure 5(a) are presented in figure 5(b) with solid black circle, a solid red star and a blue down-triangle, respectively. The black, red and blue dash lines are the expected exponents of a first-order phase transition ($d = 3$), the three-dimensional Ising universality class ($\gamma/\nu$ = 1.9630(30)) ~\cite{Ising exponent}, and crossover (0), respectively.

At $h = 0.0005$, $\lambda_{\chi_2}=2.46547 \pm 0.0786$. It is less than the expected exponent 3 here. This is consistent with the indication of Ref.~\cite{weak first scaling}. In the three-dimensional, three-state Potts model, with the increase of the external field to the critical value, the first-order phase transition becomes weaker and weaker. When the first-order phase transition is very weak, it will not exhibit an $L^d$ type of scaling in the transition region, but a second-order-type scaling~\cite{weak first scaling}.

At $h = 0.000775$, $\lambda_{\chi_2}=1.96591\pm 0.0241$. It is very close to that of the Ising universality class. The critical point of the three-dimensional, three-state Potts model belongs to the three-dimensional Ising universality class~\cite{second order 2}, where the critical behavior is controlled by two relevant scaling couplings, the temperature-like coupling ($\tau$) and field-like coupling ($\zeta$). In the vicinity of the critical point ($\beta_c$, $h_c$), there is a linear mapping to the Ising scaling couplings,
\begin{flalign}\label{linear ansats 1}
&\qquad \tau = \beta-\beta_c + a(h-h_c),& \nonumber \\
&\qquad \zeta = h-h_c + b(\beta-\beta_c),&
\end{flalign}
where the mixing parameters $a$ and $b$ are determined in Ref.~\cite{second order 2}. Although the magnetization $M$ loose its meaning as operator being conjugated to the field-like coupling $\zeta$, the exponent $\lambda_{\chi_{2}}$ of its susceptibility $\chi_2$ will also be very close to the critical exponent ratio ($\gamma/\nu$) as that in the three-dimensional Ising universality class~\cite{Karsch-plb488}.

At h = 0.002, $\lambda_{\chi_2}= 0.239149\pm 0.0054366$. It is slightly larger than 0 for crossover. There is still weak system size dependence. For crossover, it needs very large volumes to get saturation. With increase of external field, i.e., deeper into the crossover region, the saturation happens at smaller volumes. So it shows the current system sizes from $L=40$ to $L=70$ is still not large enough for size independent at $h=0.002$.

In figure 4, it is clear that the fourth order susceptibility has a negative minimum near the phase transition temperature, if the finite-size scaling of its depth $|\chi_4^{min}|$ exists, it could be as follows,
\begin{flalign}\label{fss chi4}
&\qquad |\chi_4^{min}(L)| \propto L^{\lambda_{\chi_4}}.&
\end{flalign}
Here, the scaling exponent $\lambda_{\chi_4}$ should equal to $3d$ and 0 for a first-order phase transition and crossover, respectively. For a second-order phase transition, it is generally between $3d$ and 0. The scaling exponent can also be gotten from the logarithm of equation \eqref{fss chi4}
\begin{flalign}\label{ln fss chi4}
&\qquad \ln|{\chi_4^{min}}| = \lambda_{\chi_4} \ln L + C_2,&
\end{flalign}
where $C_2$ is a constant.

Figure 5(c) shows the logarithm $\ln|\chi_4^{min}|$ as a function of $\ln L$ at various system sizes at $h = 0.0005$ (solid black circles), 0.000775 (solid red stars), and 0.002 (solid blue down-triangles). The same color points show similar linear behavior as those in figure 5(a).

The slopes of three solid lines ($\lambda_{\chi_4}$) in figure 5(c) are presented in figure 5(d) with a black circle, red star and blue down-triangle, respectively. They are 8.14359 $\pm$ 0.282178, 7.12512 $\pm$ 0.1  and 2.1397 $\pm$ 0.0048. The three dash lines from top to bottom are corresponding exponents of first-order phase transition ($3d=9$), the three-dimensional Ising universality class ($d+2\gamma/\nu$ = 6.926)~\cite{Ising exponent}, and crossover (0). The black circle, red star and blue down-triangle are between two up lines, very close to the middle line, and close to the bottom line, respectively. So the exponent $\lambda_{\chi_4}$ change gradually from less than 9 to the critical exponent ratio of the Ising universality class, and finally approaching 0 for crossover, similar to the case of susceptibility in figure 5(b).

When crossing the phase boundary at different given external fields of the three-dimensional, three-state Potts model, the maximum (minimum) of $\chi_2$ ($\chi_4$) is well described by finite-size scaling. The exponent of scaling changes gradually from 3 (9) to 0 when the value of the external field increases from zero.

\section{Summary and conclusions}

Using the three-dimensional, three-state Potts model, we study the finite-size behavior of the second to fourth order generalized susceptibilities in the vicinity of the phase transition temperatures of external fields of first-, second-order phase transitions and crossover. We find that these generalized susceptibilities have qualitatively similar features. A peak appears in the second order susceptibility, and oscillation, or a sign change happens at the third and fourth order susceptibilities. So non-monotonic fluctuations of the generalized susceptibilities are signatures of the critical point, but also can be observed at the first-order phase transition and crossover.

It is also shown that the finite-size scaling of the second and fourth order generalized susceptibilities hold at the external fields of the first-, second-order phase transitions and crossover, respectively. Exponents of its scaling change gradually from less than the value given by a first-order phase transition, to the value consistent with that of the Ising universality class, and finally approaching to zero expected for crossover.

So along the phase boundary of the three-dimensional, three-state Potts model, exponents of the finite-size scaling of the second and fourth order susceptibilities change gradually between 3 (or 9) and 0. The order of the transition can be identified by the value of the exponent of finite-size scaling of generalized susceptibilities.

\section{Acknowledgement}

This work is supported in part by the Major State Basic
Research Development Program of China under Grant No.
2014CB845402, the NSFC of China under Grants No.
10835005, No. 11221504, No. 11005046, and No. 11005045,
and the Ministry of Education of China with Project No.
20120144110001.

\end{document}